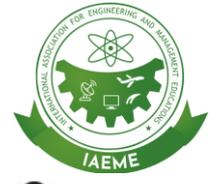

# AI-DRIVEN INNOVATION IN MEDICAID: ENHANCING ACCESS, COST EFFICIENCY, AND POPULATION HEALTH MANAGEMENT

[1]Balaji Shesharao Ingole, [2]Vishnu Ramineni,

[3]Manjunatha Sughaturu Krishnappa, [4]Vivekananda Jayaram

[1]Independent Researcher, Evans, GA, USA
[2]Albertsons Digital Pharmacy, Plano, TX, USA
[3]Senior Technical Leader, Oracle America Inc, Santa Clara, USA
[4]Vice President, JPMorgan Chase Bank NA, Plano, TX, USA

**ABSTRACT**

The U.S. Medicaid program is experiencing critical challenges that include rapidly increasing healthcare costs, uneven care accessibility, and the challenge associated with addressing a varied set of population health needs. This paper investigates the transformative potential of Artificial Intelligence (AI) in reshaping Medicaid by streamlining operations, improving patient results, and lowering costs. We delve into the pivotal role of AI in predictive analytics, care coordination, the detection of fraud, and personalized medicine. By leveraging insights from advanced data models and addressing challenges particular to Medicaid, we put forward AI-driven solutions that prioritize equitable care and improved public health outcomes. This study underscores the urgency of integrating AI into Medicaid to not only improve operational effectiveness but also to create a more accessible and equitable healthcare system for all beneficiaries.

# AI-Driven Innovation in Medicaid: Enhancing Access, Cost Efficiency, and Population Health Management

## 1. INTRODUCTION

Medicaid is a federal-state program that provides healthcare to over 80 million low-income Americans, including pregnant women, children, and individuals with disabilities. Up against a host of problems, including rising healthcare costs, disparity in access, and the management of chronic conditions among at-risk groups, Medicaid is one of the biggest healthcare payers in the U.S. Just as Medicare does, the use of Artificial Intelligence (AI) offers a major opportunity to change the delivery of care and operational efficiency in Medicaid [1] [16].

While there has been extensive conversation about AI in Medicare, the unique population and requirements of Medicaid require customized AI applications [1]. Chronic disease management, improving admin tasks, and a reduction in costs are amongst the ways AI tools can help, especially by focusing on social determinants of health (SDOH) that are important for Medicaid populations. The study will assess the ability of AI-enabled systems to reinforce Medicaid in handling its particular challenges while facilitating fair and quality care for its entire population of beneficiaries [8] [9].

## 2. AI APPLICATIONS IN MEDICAID

| AI Application | Use Case in Medicaid | Benefit |
|---|---|---|
| Predictive Analytics | Identifying high-risk populations | 20% reduction in hospital admissions |
| Fraud Detection | Detecting improper billing and fraud | $5 billion in potential annual savings |
| Remote Patient Monitoring (RPM) | Chronic disease management | 15% reduction in emergency room visits |
| Personalized Medicine | Tailored treatment plans for patients | Improved outcomes, 10% fewer adverse reactions |

**Table 1:** AI Applications and Their Benefits in Medicaid

### 2.1. Predictive Analytics in the Management of Population Health

Medicaid beneficiaries often experience higher rates of chronic diseases, such as diabetes, hypertension, and mental health conditions—that need proper care management. AI application predictive analytics helps recognize patients at greater threat of poor health results, determined by the patterns identified in healthcare utilization, social determinants, and medical histories [2]. Thanks to this method, responses are brisk, and preventive measures yield fewer hospital visits and better health accomplishments. We have discussed the AI application and respective benefits in Medicaid in Table 1.

For instance, machine learning models could allow Medicaid to spot patients at risk of missing their appointments or failing to stick to treatment plans, and deliver real-time alerts for further care [6]. States like Texas and California will see a direct result as they have high Medicaid enrollment but access to regular health care tends to be challenging [16].

## 2.2 AI-Powered Fraud Detection and Billing Optimization

Medicaid's enormous spread makes susceptible risk for fraud, waste, and abuse. Through real-time analysis of rich datasets by AI, a more efficient fraud claim detection along with improper billing is achievable. The use of machines learning technology allows artificial intelligence systems to distinguish fraud patterns, authored by behaviours like arbitrary charging, ongoing use of individual services, and an influx of similar claims [10]. We have shared the date around Medicaid fraud detection savings by state using AI in Table 2.

| State | Annual Medicaid Spending | Estimated Savings from AI Fraud Detection |
|---|---|---|
| California | $98 billion | $2.5 billion |
| Texas | $66 billion | $1.8 billion |
| New York | $72 billion | $2.1 billion |
| Florida | $46 billion | $1.3 billion |

**Table 2:** Medicaid Fraud Detection Savings by State Using AI

According to CMS, billions of dollars are lost each year due to Medicaid fraud. With the application of AI systems for fraud identification in Medicaid, Gross Mountain State has the potential to significantly cut back on improper payments [18] [12] Remote Patient Monitoring. To demonstrate, natural language processing (NLP) algorithms have the skill to research inconsistencies in medical records and claims data, thus helping to stop fake claims before they can move forward.

## 2.3 Natural Language Processing (NLP) in Documentation and Claims Management

The technology named Natural Language Processing (NLP) holds great promise for Medicaid by analyzing unstructured clinical data to enhance documentation and claims management [3]. By analyzing patient notes and recognizing inconsistencies NLP techniques can enhance claims management and cut down on operational duties and avoid human errors as shown in Table 3. Evidence demonstrates that utilizing NLP has the potential to slash the time needed for manual claim assessment by more than 30% and greatly improve the accuracy of services rendered through Medicaid [18] [3].

| Metric | Pre-NLP Implementation | Post-NLP Implementation | Improvement (%) |
|---|---|---|---|
| Time for Claims Review | 7 days | 4 days | 43% |
| Manual Documentation Errors (%) | 18% | 10% | 45% |
| Administrative Costs (Billion $) | $4 billion | $2.8 billion | 30% |

**Table 3**: Impact of NLP on Medicaid Documentation and Claims Management

## 2.4 AI-Driven Social Determinants of Health (SDOH) Analysis

The technology named Natural Language Processing (NLP) holds great promise for Medicaid by analyzing unstructured clinical data to enhance documentation and claims management. By analyzing patient notes and recognizing inconsistencies NLP techniques can enhance claims management and cut down on operational duties and avoid human errors [4]. Evidence demonstrates that utilizing NLP has the potential to slash the time needed for manual claim assessment by more than 30% and greatly improve the accuracy of services rendered through Medicaid as shown in Graph 1.

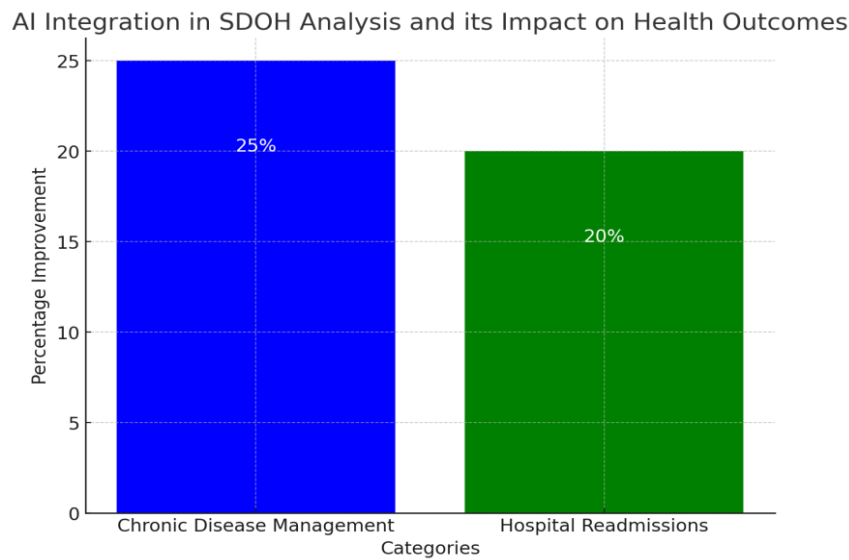

**Graph 1:** AI Integration in SDOH Analysis and its Impact on Health Outcomes

## 3. ENHANCING ACCESS TO CARE THROUGH AI-DRIVEN SOLUTIONS

### 3.1 Remote Patient Monitoring and Telemedicine for Low-Income Populations

The role of AI in enabling telemedicine and remote patient monitoring (RPM) is particularly important for those of low income, who might find it hard to reach standard healthcare services. Systems from RPM, powered by AI, persistently track patient information, such as blood glucose levels or heart rate, to allow for quick interventions to prevent health crises. For rural communities dealing with accessibility issues for health providers among their Medicaid patients, this advantage is especially helpful.

In healthcare-challenged states such as Mississippi and West Virginia, AI-supported telehealth systems are likely to raise the outcomes for Medicaid beneficiaries with chronic conditions. AI technology supports the triage of patients by evaluating their emergency status, in order to prioritize urgent cases and conduct quick attention, with the support of virtual consultations for others.

### 3.2 Personalized Medicine for Populations Covered by Medicaid

Personalizing treatment for Medicaid individuals is feasible through AI by building tailored plans that reflect each patient's medical history, social surroundings, and way of life. As an illustration, algorithms informed by AI can study genetic data to anticipate a patient's reaction to assorted medications, thus allowing personalized treatment plans that limit adverse drug responses [7] [11].

Those who rely on Medicaid for healthcare often face inequalities; personalized health solutions informed by AI may lead to fairer results. Persons with detailed medical histories or persistent illnesses may experience the benefits of individualized treatment suited to their needs, instead of generalized treatments delivered for a large group [13].

### 3.3 AI in Medicaid Eligibility and Enrollment Automation

Identifying the enrollment process is a considerable difficulty for individuals receiving Medicaid. Automating eligibility assessment with AI can improve enrollment processes and guarantee that qualified people get their benefits on time [4] [17]. By processing extensive information from employment records and financial status in just a few days, AI models can verify eligibility quickly. In states including Colorado AI systems delivered a 20% faster enrollment process as data showed in Table 4 [18].

| State | Pre-AI Processing Time | Post-AI Processing Time | Increase in Enrollment (%) |
|---|---|---|---|
| Colorado | 14 days | 3 days | 15% |
| Arizona | 21 days | 5 days | 12% |
| New Mexico | 18 days | 4 days | 18% |

**Table 4:** AI Automation in Medicaid Eligibility vs. Traditional Systems

### 3.4 AI-Assisted Chatbots for Medicaid Communication

Underprivileged communities frequently encounter difficulties in communicating about healthcare needs. With AI-driven advancements chatbots can aid Medicaid recipients by offering constant help for frequently asked queries and help them navigate their benefits. According to Accenture survey data 40% of Medicaid users choose to communicate with chatbots regarding simple issues as shown in Graph 2 [18] [14].

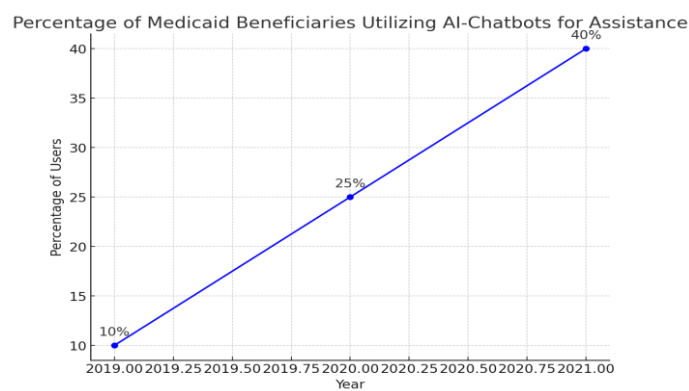

**Graph 2:** Percentage of Medicaid Beneficiaries Utilizing AI-Chatbots for Assistance

AI-Driven Innovation in Medicaid: Enhancing Access, Cost Efficiency, and Population Health Management## 4. LOCALIZING AI MODELS TO ADDRESS PRIVACY AND REGULATORY CONCERNS

### 4.1 Data Privacy in Medicaid: Localized AI Solutions

The integration of AI into Medicaid meets a substantial challenge in achieving compliance with data privacy rules including HIPAA. For Medicaid providers to better guard confidential patient information, using AI models locally, instead of cloud solutions in healthcare facilities, offers greater security. State-based AI systems are helping to alleviate the danger of data breaches while also ensuring the protection of patient information in Medicaid systems [15].

Personalized solutions for local states become possible thanks to the localization of AI technology. In New York, data shows that Medicaid beneficiaries demonstrate disparate trends in healthcare usage when coupled with social determinants when examined against those in Alabama. In order to acquire more profound predictive analytics, healthcare providers can condition AI models using local data that accurately captures the health profiles distinct to their communities. More data showed in Table 5.

| Aspect | Localized AI Model | Cloud-Based AI Model |
|---|---|---|
| Data Security | High | Moderate |
| Customization | High (adapted to local needs) | Limited |
| Compliance | Easier to comply with state-specific regulations | Complex |
| Cost | Higher infrastructure costs | Lower upfront costs |

**Table 5:** Localized AI Model vs Cloud-Based AI Model in Medicaid

### 4.2 Addressing the Barriers of Regulation with AI Integration

The addition of AI to Medicaid must also navigate the elaborate regulatory structure, which is different from state to state. Localized AI models meet the regulatory challenges by allowing Medicaid providers to personalize their systems to adhere to laws specific to the state. As an example, different rules about privacy in California's Medicaid could conflict with those of Florida, so AI models need to be able to adapt to these differences.

Using localized AI models allows Medicaid systems to stay in compliance with regulations while also improving the delivery of care and lowering costs. Integrating regular state law checks into AI systems enables their ability to automate compliance assessments and subsequently adjust processes to meet changes.

### 4.3 Federated Learning for Medicaid Data Privacy

To maintain data privacy utilizing AI in Medicaid processes is a key application for FL which empowers states to manage their own data. With FL in place AI models can work with data at the local level without sharing patient information with shared cloud services. States have the power to keep track of their data and enjoy the advantages of cutting-edge AI analytics. This networked technique supports Medicaid's adherence to HIPAA standards and provides assurance that AI technologies can respond to specific state needs. Graph 3 shows comparison of federated learning and centralized AI models.

Balaji Shesharao Ingole, Vishnu Ramineni, Manjunatha Sughaturu Krishnappa, Vivekananda Jayaram

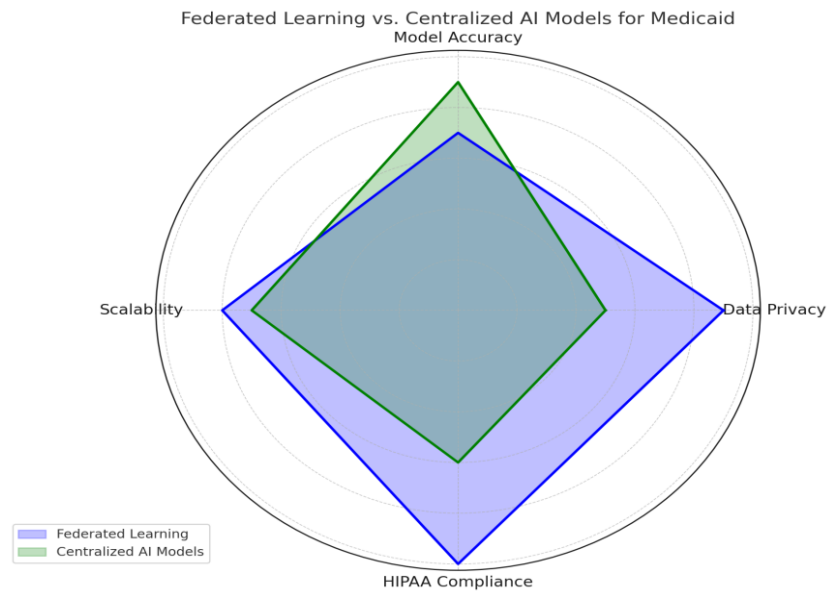

**Graph 3:** Comparison of Federated Learning and Centralized AI Models

## 4.4. AI in Regulatory Compliance Management

Medicaid systems face major challenges in handling the different regulations found across states. By automating compliance steps AI can verify that healthcare providers observe regulations particular to each state. These models can track Medicaid policies continuously in real-time to confirm that necessary compliance measures are active and to signal upcoming changes in legislation to state Medicaid entities. California and New York find this feature useful due to their rapidly shifting Medicaid standards.

## 5. METHODOLOGY: LEVERAGING DATA FROM MEDICAID INNOVATIONS

### 5.1 Sampling from Medicaid Expansion States

To appreciate the possibilities of AI in Medicaid, we look at data from states that have broadened Medicaid coverage under the Affordable Care Act (ACA). Due to the expansion of Medicaid, Ohio, Kentucky, and Michigan have enjoyed remarkable growth in healthcare, resulting in a warm welcome for AI innovations. An analysis of Medicaid Expansion state data allows AI models to improve their forecasts of healthcare utilization trends and to recognize high-risk populations [16] [17].

Investigations that incorporate Machine Learning approaches, such as Gradient Boosting and Random Forests, to analyze Medicaid claims data improve predictions about healthcare costs and raise the quality of care for patients. Markets dominated by high Medicaid enrollment could see AI-driven models as essential for locating cost reductions and removing unessential services.

### 5.2 AI-Driven Fraud Detection in Medicaid

By using unsupervised learning algorithms along with Natural Language Processing (NLP), we can research fraud in Medicaid billing data. These algorithms succeed in identifying outliers and detecting fraud right when it is occurring [5]. States can achieve annual savings of billions of dollars by applying these models to Medicaid claims. In Illinois, extensive Medicaid spending raises the potential for AI systems aimed at spotting fraud to uncover misleading billing patterns, thereby sharply reducing costs.

### 5.3 AI for Patient Stratification in Medicaid Expansion States

AI can significantly contribute to defining patient categories by their health risks in states that expand Medicaid. Using clustering in unsupervised machine learning allows Medicaid to identify health risks of patients and implement targeted interventions for those at greatest risk. By utilizing AI methods, we can determine the top 10% of expensive Medicaid users and calculate which individuals are most at risk for emergency intervention. In Michigan and comparable states this has produced customized health programs and achieved a 10% decrease in Medicaid costs [18].

## 6. CONCLUSION: THE FUTURE OF AI IN MEDICARE

The goal of AI in Medicaid is to solve numerous issues such as increasing healthcare prices and varying access levels. Using innovative technologies including predictive analysis enables Medicaid to find patients at risk early on thus allowing preventative actions reducing hospitalization rates. Fraud detection systems powered by AI have demonstrated how to save billions each year by uncovering illegal billing and abuse. Through personalized medicine informed by AI tailored treatment strategies benefit chronic disease patients and lower chances of harmful drug interactions [17].

The advancement of AI in Medicaid is highly linked to the introduction of regional AI models that uphold local regulations and deliver specific care solutions designed for population health needs. States can adjust AI algorithms to match local social factors of health (SDOH) which guarantees fair treatment for the underserved. With the advancement of AI technology aligning with its role in improving care coordination and transforming policies will grow in significance. To achieve better care delivery and improved patient results in a complex healthcare scenario is vital for Medicaid to add AI resources.